\documentclass[twocolumn,pre,showpacs,amsmath,amssymb]{revtex4}
\pdfoutput=1
\usepackage{epsfig}
\usepackage{graphicx}
\usepackage{dcolumn}
\usepackage{bm}

\begin{document}

\title{Subcritical instabilities in a convective fluid layer under a quasi-1D heating}

\author{M.A. Miranda}
\affiliation{Dept. of Physics and Applied Mathematics, Universidad de Navarra. Irunlarrea s/n, E-31080 Pamplona, Spain}

\author{J. Burguete}
\email[e-mail address:\ ]{javier@fisica.unav.es}
\affiliation{Dept. of Physics and Applied Mathematics, Universidad de Navarra. Irunlarrea s/n, E-31080 Pamplona, Spain}

\date{\today}

\begin{abstract}
The study and characterization of the diversity of spatiotemporal patterns generated when a rectangular layer of fluid is locally heated beneath its free surface is presented. We focus on the instability of a stationary cellular pattern of wave number $k_s$ which undergoes a globally subcritical transition to traveling waves by parity-breaking symmetry. The experimental results show how the emerging traveling mode ($2/3k_{s}$) switches on a resonant triad ($k_s$, $k_s/2$, $2k_{s}/3$) within the cellular pattern yielding a ``mixed'' pattern. The nature of this transition is described quantitatively in terms of the evolution of the fundamental modes by complex demodulation techniques. The B\' enard-Marangoni convection accounts for the different dynamics depending on the depth of the fluid layer and on the vertical temperature difference. The existence of a hysteresis cycle has been evaluated quantitatively. When the bifurcation to traveling waves is measured in the vicinity of the codimension-2 bifurcation point, we measure a decrease of the subcritical interval in which the traveling mode becomes unstable. From the traveling wave state the system undergoes a {\it new} global secondary bifurcation to an alternating pattern which doubles the wavelength ($k_{s}/2$) of the primary cellular pattern, this result compares well with theoretical predictions [P. Coullet and G. Ioss, {\em Phys. Rev. Lett.} {\bf 64}, 866 (1990)]. In this cascade of bifurcations towards a defect dynamics, bistability due to the subcritical behavior of our system is the reason for the coexistence of two different modulated patterns connected by a front. These fronts are stationary for a finite interval of the control parameters.
\end{abstract}


\maketitle

\section{Introduction}
\label{intro}

The experimental and theoretical study of the dynamics of 1D cellular patterns is still a subject of great interest, specially when this research is devoted to spatially extended systems. The simplicity of such systems, combined with the theoretical approach by nonlinear dynamics and critical phenomena, allows us to discover universal mechanisms at the characteristic length scales without being necessary to cope with the complex microscopic problem.
These 1D dynamics for spatially extended dissipative systems have been reported for a large variety of experiments where the bifurcation scenario from a homogeneous pattern can be a {\it primary bifurcation} either to a stationary cellular pattern or to a traveling wave pattern. As the control parameter is increased the system may go through a {\it secondary bifurcation}.
Some of these experiments are:  the printer's instability~\cite{Pan94,Giorgiutti95}, circular liquid column arrays~\cite{Brunet00,Mazel99,Counillon98}, directional solidification~\cite{Simon88}, directional viscous fingering~\cite{Bellon98}, the Taylor-Dean system~\cite{Bot98}, electroconvection in liquid crystals with rectangular cell~\cite{Pastur01,Joets89}, lateral heating in a rectangular cell~\cite{Burguete01,Burguete99,Garnier03,Pelacho99,Pelacho00}, lateral heating in an annular cell~\cite{Daviaud92}, 1D heating in a rectangular cell~\cite{Burguete03,Burguete93,Maza94,Alvarez97,Ringuet93,Ringuet99,Gouesbet00,Pastur03}, numerical simulations with the stabilized Kuramoto-Sivashinsky equation~\cite{Misbah94}. \par 

 The aim of this paper is the study of a dissipative extended system, a buoyant-thermo-capillary driven flow, which undergoes a cascade of secondary bifurcations from a stationary cellular state towards a defect dynamics. Certain symmetries are broken in each bifurcation. Specifically, we study experimentally the mode competition at the threshold of a secondary instability to traveling waves. This convective approach allows us to carry out experimental measurements where the relevant magnitudes can be easily measured, with the cost of relatively large time-scales required to achieve asymptotic states. \par
 Our system consists of a thin rectangular fluid layer of silicone oil opened to the atmosphere with an inhomogeneous heating at the center and along a line placed at the bottom.   
This experimental set-up is intended to approach an ``ideal'' quasi-1D thermoconvective system. The fluid dynamics is in the context of the Oberbeck-Boussinesq approximation for the applied range of temperatures. The physically relevant control parameters are the vertical temperature difference $\Delta T_v$ and the depth of the fluid layer $d$.\par

Without threshold, as soon as $\Delta T_v \not=$ 0, a homogeneous convective pattern (PC) appears: the fluid over the heater line (HL) raises by the effect of the negative vertical temperature gradient and it returns down near the cooling walls. Therefore, a circulating flow at both sides of HL develops the structure of two counterotating rolls whose axis are parallel to the HL. As we increase the control parameter $\Delta T_v$ from PC, this pattern will invariably remain for the subsequent instabilities. For asymptotic time responses we can distinguish three fundamental patterns: a stationary cellular pattern (ST) with wave number $k_s$, a traveling pattern (TW) consisting of one propagative wave with wavenumber $2k_s/3$ and an alternating pattern (ALT) consisting of two traveling counterpropagative waves with wavenumbers $k_s/2$ plus the stationary one (a resonant triad). Similar patterns have been analyzed in a previous experimental setup~\cite{Burguete03}, but with a less developed system. In this work, accuracy has been improved by means of new optical and data acquisition systems.\par

In 1990, Coullet and Ioss~\cite{Coullet90} (CI) classified the global character of secondary bifurcations in 1D systems using a Bloch-Floquet analysis; and predicted ten generic secondary instabilities. In this theory, there is a coupling between the phase of the basic pattern and the amplitude of the unstable modes associated to the secondary bifurcation. 
On the other hand, the phase-amplitude coupled equations obtained with less restrictive symmetry arguments by L. Gil~\cite{Gil99,Gil00} are suitable for experiments with analogous dynamics, but six real and fifteen complex coefficients are necessary to be determined experimentally.\par  
  
We report the instability of a ``mixed'' pattern due to the close position of the new emerging mode of the TW. This mixed pattern consists of localized ALT domains with fluctuating boundaries over ST. These results shed new light on the role of two close competing oscillatory modes in the onset of a subcritical transition to TW. These are the emerging traveling mode and the nearest oscillatory one with wavenumbers $2k_s/3$ and $k_s/2$, respectively. This relationship dos not fit into the classification made by CI, and this fact could be the consequence of ({\it i}) subcriticality or ({\it ii}) the dynamical implication of the other two spatial directions.\par
 
In the bistable regimes for two subsequent subcritical bifurcations to TW and ALT, we report existence of {\it normal fronts}~\cite{Pomeau86} that connect two stable states of the oscillatory type. These fronts are stationary and accordingly to the results reported here, they are pinned to both modulated patterns for a finite control parameters range. A similar front dynamics linked to subcriticality has already been modelized for binary fluids~\cite{Bensimon88}, however in this case pinning is produced between a homogeneus and a traveling wave pattern. In transients, when we place our system below and close to the transition point, we report the study of the so called {\it Fisher-Kolmogorov-Petrovsky-Piskunov (FKPP) fronts}~\cite{Fisher37} that connect stable states with unstable states. These fronts provide interesting features of the dynamics such as the {\it convective vs. absolute} character of an instability.\par

From the theoretical point of view subcritical transitions are predicted in superconductors and in liquid crystals~\cite{Halperin74}. Despite subcriticality has been widely studied experimentally in other systems, some examples (not pretending to be an exhaustive list) are: pearling of tubular membranes~\cite{Bar94}, the Fr\'eedericksz transition~\cite{Residori04} and the transition to traveling waves in binary fluids~\cite{LaPorta98}, here we report quantitative measurements for a new subcritical transition in a quasi-1D convective system from a cellular pattern to traveling waves and further beyond. \par

\section{EXPERIMENTAL SETUP}
\label{sec:setup}

\begin{figure}
\includegraphics[width=8cm]{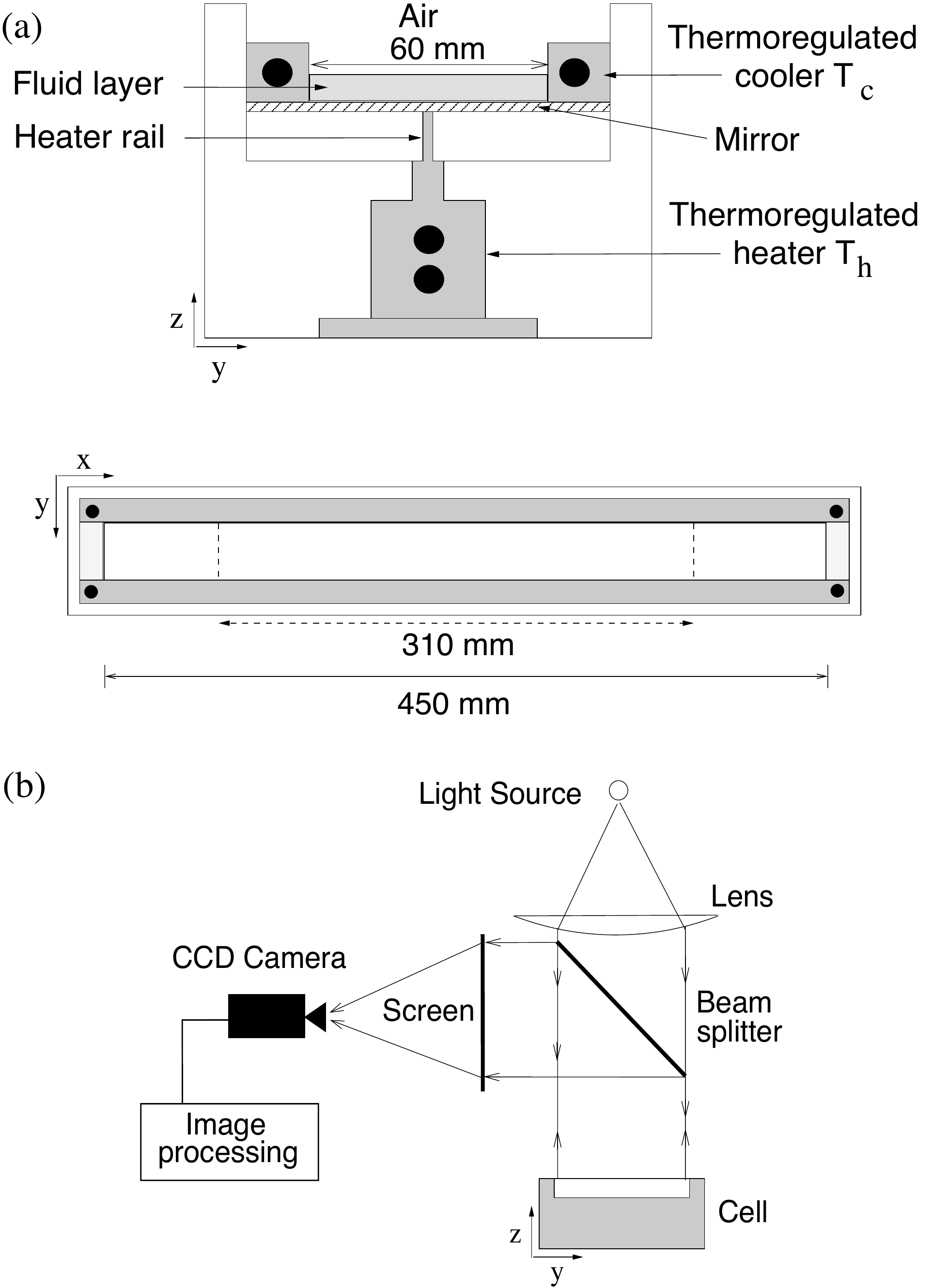}
\caption{\label{fig:1}Sketch of the experimental setup. (a) Rectangular cell: cross section (up) and top view (bottom) of the rectangular vessel. The scaling in both views is different; (b) Optical setup.}
\end{figure}

The fluid layer of depth $d$ is placed in a narrow convective channel $L_x \times L_y$ ($L_x = $ 470 mm, $L_y = $ 60 mm). As shown in Fig.~\ref{fig:1}a this layer lies over a flat surface, a mirror of thickness 3 mm. The bottom of the mirror is in contact with a heater rail of 1 mm (thickness) which provides the heating line in the $\hat x$ direction, HL. The heater rail belongs to an aluminum block with an inner closed water circuit thermoregulated by a heater bath at $T_h$. The temperature measured at HL on the mirror surface differs from the one given by the heater bath probe in $-3.00$ $^\circ$C for $d=$ 4 mm and $T_h=$ 48 $^\circ$C. We limit the heat transfer in the $\hat y$ direction by isolating laterally the heater rail with Plexiglas. Therefore in the transversal direction of the cell, $\hat y$, one may approximately achieve a Gaussian temperature profile smoothed nearby the central heater because of the thermal diffusivity of the mirror. Regarding other experimental setups like the resistive wire experiments (20 $\mu$m - 50 $\mu$m diameter)~\cite{Ringuet93,Ringuet99,Gouesbet00} where there is a coupling between the circulation flow and the thermal stability of the wire, in our system because of the great thermal inertia couplings between the convection in the bulk and the heater are not expected. The lateral walls are two aluminum blocks (coolers) whose temperatures are kept constant at $T_c = $ 20 $\pm$ 0.1 $^\circ$C by means of a secondary water circulation. This temperature is given by the cooler bath probe and it is the same as the one measured at the top of the aluminum blocks. The shorter boundary walls at the opposite extremes on $\hat x$ are made of Plexiglas. All these experimental parts are contained in a Delrin block. The effective dimensions of the cell are $450$ mm $\times$ 60 mm $\times$ 15 mm ($L_x \times L_y \times L_z$), although the optical setup allows us to visualize only a centered area from an upper surface of $310$ mm $\times$ 60 mm. \par
The geometric aspect ratios for $d$ = 4 mm are: $\Gamma_x=L_x/d =$ 112.5, $\Gamma_y=L_y/d =$ 15. The ratio $\Gamma_x/\Gamma_y=$ 7.5 allows us to classify the system as weakly confined in $\hat x$.\par

The fluid used is a silicone oil (brand name: {\it Dow Corning 200}) with a nominal viscosity of 5 cSt (see properties on Table~\ref{tab:1}). This fluid is transparent to visible light and therefore it is suitable for the shadowgraph technique. It has a low volatility, the depth of the fluid layer diminishes with a rate of 0.01 mm/24 hours over an area of 270 cm$^2$ at 20 $^\circ$C, so we can work (in the range of temperatures applied) with the layer opened to the atmosphere.  Under this condition, we assume that the depth, measured with a precision of 0.01 mm with a micrometric screw, remains constant along each measurement. The temperature outside the cell is controlled at $T_a = $ 20 $\pm$ 0.1 $^\circ$C for a sequence of measurements.\par
From the corresponding physical properties (Table~\ref{tab:1}) there are two significant adimensional numbers: the Prandtl number $Pr=\nu/\kappa \approx 75$ ($\equiv$ compares {\sl viscous diffusivity} $vs.$ {\sl thermal diffusivity}) and the dynamic Bond number $Bo_D = R/M \approx$ 1.05 ($d=$ 3 mm) ($\equiv$ compares {\sl thermogravitatory effects} $vs.$ {\sl thermocapilarity effects}). As well at $d =$ 3 mm and $\Delta T_v =$ 30 K, we can determine two characteristic time scales: the viscosity time scale $\tau_\nu =$ 2 s and the thermal diffusivity time scale $\tau_\kappa =$ 135 s, thus the dynamics is driven by the temperature field.

\begin{table}
\caption{\label{tab:1}Physical properties of the 5 cSt silicone oil.}
\begin{ruledtabular}
\begin{tabular}{lllllll}
\noalign{\smallskip}
Surface tension (25$^\circ$C) & $\sigma = $ 19.7 mN m$^{-1}$\\
Thermal conductivity (50$^\circ$C) & $\lambda = $ 0.117 W m$^{-1}$ K$^{-1}$\\ 
Thermal diffusivity $(\lambda/\rho c_p)$ & $\kappa = $ 6.68$\: 10^{-8}$ m$^{2}$ s$^{-1}$\\
Kinematic viscosity (25$^\circ$C) & $\nu = $ 5 cSt (5 10$^{-6}$ m$^{2}$ s$^{-1}$)\\
Density (25$^\circ$C) & $\rho = $ 913 kg m$^{-3}$\\
Refractive index (25$^\circ$C) & $n = $ 1.3960\\
Linear expansion coefficient & $\alpha = $ 0.00105 K$^{-1}$\\
Surface tension/temperature & $\frac{\partial\sigma}{\partial T} = $ -8 10$^{-5}$ N m$^{-1}$ K$^{-1}$\\
\noalign{\smallskip}
\end{tabular}
\end{ruledtabular}
\end{table}

The shadowgraphic flow-visualization system is sketched in Fig.~\ref{fig:1}b. In order to obtain quantitative results for a sequence of measurements, the light intensity from an incandescent light bulb (white and incoherent light) is controlled with a power supply.  An incoming parallel light is sent through the convection pattern. The light beam crossing the fluid layer suffers different deflections depending on the refractive index gradients (i.e. temperature gradients). The analysis of the spatiotemporal periodicity of the modulation of the light beam, that has crossed the fluid layer, provides information about the temperature field of a certain pattern. In a similar experimental cell the deflection induced by traveling waves was below 1 $\mu$m for a depth of 3 mm~\cite{Burguete03}, therefore the information analyzed comes from the convective flow in the bulk. Once it is reflected back at the mirrored bottom the output beam is projected into a screen. The shadowgraph method allows us to observe the thermal gradients in the bulk of the fluid layer as a modulation of the light intensity on the screen.
The screen image is recorded with a CCD video camera with a resolution of 570 px $\times$ 485 px (px $\equiv$ pixels), keeping constant the focal length, the aperture and the gain. 
 The shadowgraphy image is an instantaneous image of the screen (Fig.~\ref{fig:2}). There, we can observe a central dark line (HL) and a bright modulation in both sides of HL by symmetry due to the effect of thermal lenses. Periodically, a line over the bright modulation next to HL is recorded to obtain a spatiotemporal diagram $S(x,t)$, and another line placed perpendicular to HL is also recorded to obtain a spatiotemporal diagram $S(y,t)$. With the developed image analysis software, the signal $S(y,t)$ is averaged over a temporal sequence to determine a medium contrasted position in $\hat y$ to place the acquisition line that records the spatiotemporal diagram $S(x,t)$ (with an acquisition rate of 1 s$^{-1}$).
The signal $S(x,t)$ is processed via a bidimensional Fourier transform and complex demodulation techniques in order to determine the amplitudes $A_{j}$, the wave numbers $k_j$ and the frequencies $\omega_j$ of each fundamental Fourier mode $M(k_j,\omega_j)$~\cite{Kolodner90}. $S(x,t)$ can also be filtered by discriminating the normalized amplitude below a critical value $\mu$.\par
The control parameters are the depth of fluid layer $d$ and the vertical temperature difference, $\Delta T_v = T_h - T_a$. Under these conditions, we have obtained the smaller threshold for the primary bifurcation at $d =$ 4.5 mm. To build the stability diagram we have explored depths $d =$ 2.5, 3, 3.5, 4, 4.5 and 5 mm. To locate the different asymptotic regimes we have worked with step $|\Delta T_h|=$ 2 K, except at $d =$ 2.5 with a bigger step ($|\Delta T_h|=$ 5 K) and some isolated measurements at $d =$ 3.5 looking for TW. Meanwhile boundaries have been defined with a maximum step of $|\Delta T_h|=$ 1 K. To determine the nature of the bifurcations at $d =$ 4 and 4.5 mm, we increase and decrease $\Delta T_v$ by steps $|\Delta T_h|=$ 1, 0.5 and 0.3 K, allowing the system to achieve the asymptotic state for 1 hour in the ascending sequences, and 3 hours in the descending sequences nearby the threshold. Experimentally, we have determined that the relaxing time for a step of $\Delta T_h=$ 1 K is approximately 1h and the relaxing time for a step of $\Delta T_h=$ -1 K is approximately 2 hours and 30 min.\par

\section{RESULTS}
\label{sec:results}

\subsection{Dynamics of the hotspots}

\begin{figure}
\includegraphics[width=8cm]{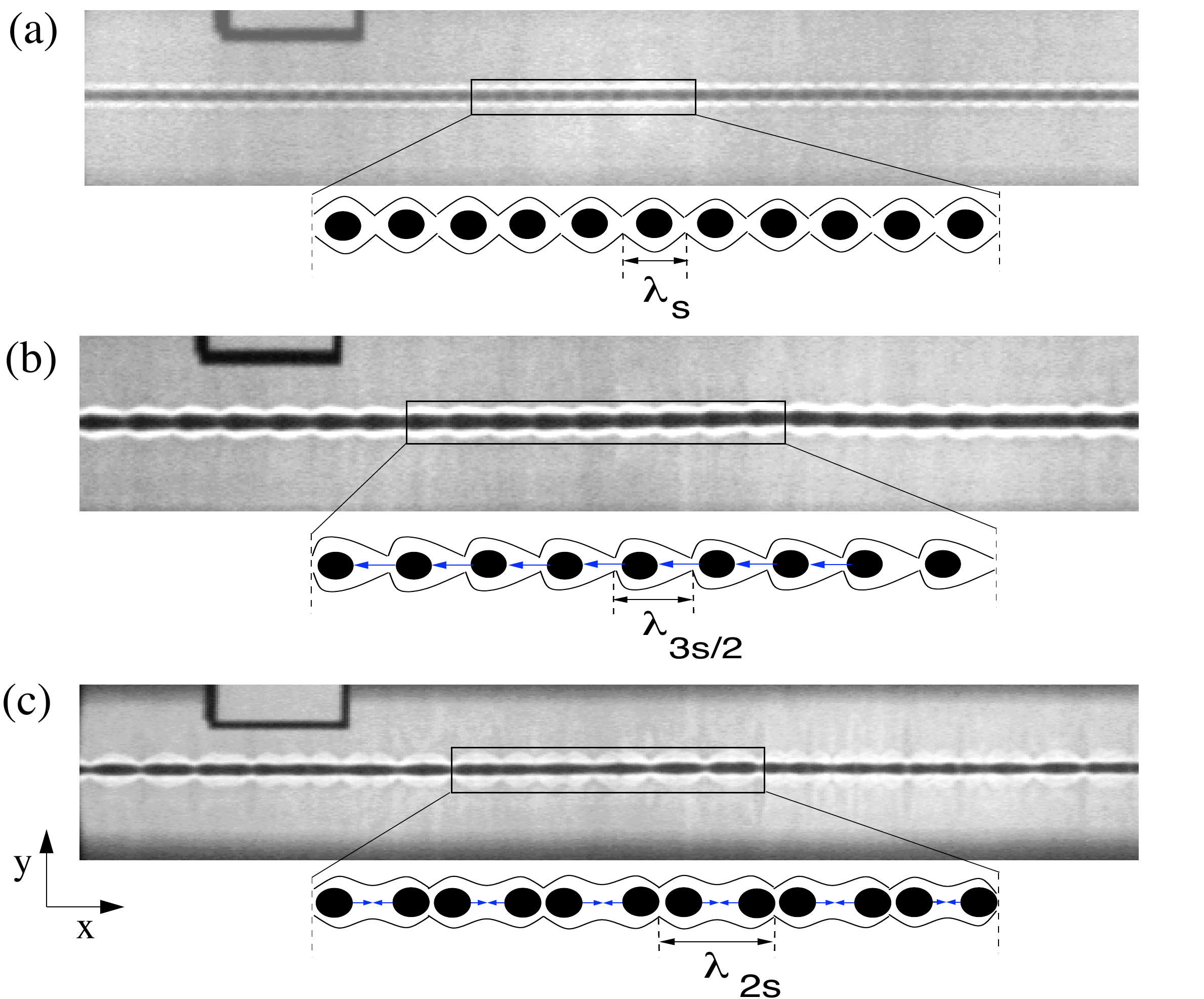}
\caption{\label{fig:2} Shadowgraphy images for three different regimes, below the dynamics of hotspots over HL is sketched (a zoom view of the selected area above): (a) stationary state ST for $d =$ 3 mm; (b) traveling waves TW for $d =$ 4 mm; (c) alternating waves ALT for $d =$ 4 mm. The arrows between hotspots indicate the direction of their movement.}
\end{figure}

The effect of thermal lenses produces on the screen a convergence of the emerging rays corresponding to the bright profiles (see Fig.~\ref{fig:2}), as well as a divergence for the ascending convective flow over HL which corresponds to a distribution of aligned dark spots or {\it hotspots}.
In Fig.~\ref{fig:2} we show the shadowgraphy images belonging to the previously introduced regimes: ST (stationary), TW (traveling wave) and ALT (alternating). Below the shadowgraphy images we have sketched how the dynamics of the hotspots is observed over HL and the space symmetries of the contours of maximum brightness.\par

If we take the wavelength of the basic cellular pattern ST, $\lambda_s$, as the space occupied by a hotspot, we have got roughly 75 oscillators which are differently coupled, regarding the type of pattern, in a dissipative system. At the same time, a driving force is applied into them when we increase $\Delta T_v$. The ratio between wavelengths of different patterns is conserved for different depths (i.e. see Fig.~\ref{fig:8}a for $d=$ 4 mm). The frequencies are shown in Fig.~\ref{fig:8}b, and the phase velocities can be obtained as $v_{\phi}=\omega/k$.\par
\begin{figure}
\includegraphics[width=8cm]{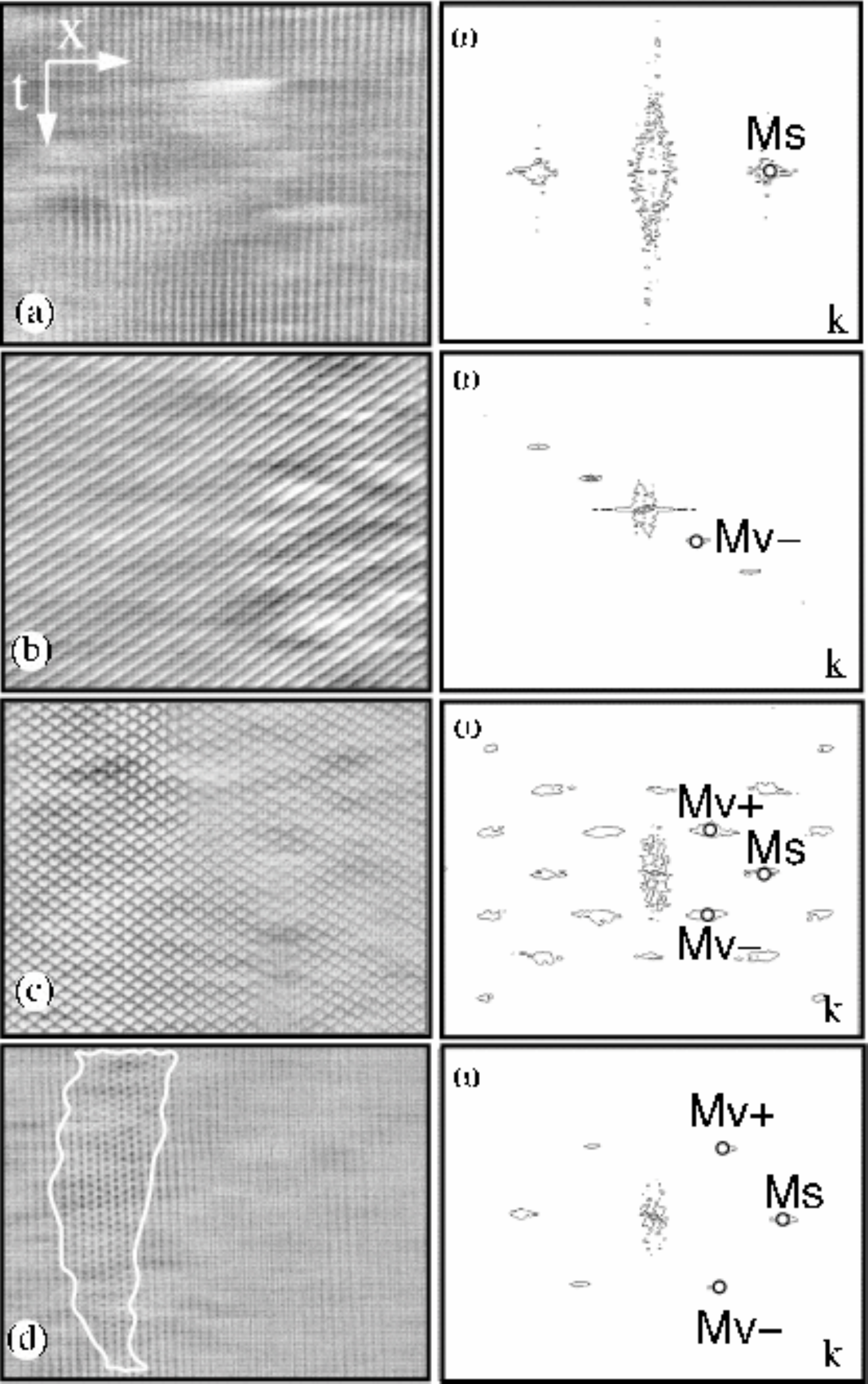}
\caption{\label{fig:3}On the left the most representative spatiotemporal diagrams $S(x,t)$, and on the right the contour plot of the corresponding Fourier spectrum $(k,\omega)$ (the fundamental modes $M_s$ (stationary) and $M_{v\pm}$ (traveling) have been pointed out, the central spot corresponds to $\left(k=0,\omega=0\right)$): (a) ST ($d =$ 3 mm, $\Delta T_v=$ 16.7 K) ; (b) TW ($d =$ 4 mm, $\Delta T_v=$ 18.6 K); (c) ALT ($d =$ 4 mm, $\Delta T_v=$ 25 K); (d) mixed pattern ST/ALT ($d =$ 3 mm, $\Delta T_v=$ 30 K), the boundary of the localized pattern ALT has been highlighted (filtering with $\mu=$ 0.37).}
\end{figure}

 
For the secondary instabilities that are going to be quantified in the following sections we show in Fig.~\ref{fig:3} the characteristic spatiotemporal diagrams $S(x,t)$  and their respective bidimensional Fourier spectra $(k,\omega)$. The fundamental modes for each pattern have been pointed out as $M_s$ (the stationary mode) and $M_{v\pm}$ (the traveling modes to the right $(+)$ and left $(-)$). The dynamics of the hotspots allow us to describe the different regimes:

\begin{itemize}
\item[(i)] The continuous spatial translation symmetry of the pattern PC is broken into a steady cellular pattern ST (Fig.~\ref{fig:3}a). In this pattern the hotspots keep their positions fixed at a distance $\lambda_s\approx 2d$ (i.e. at $d=$ 4mm see Fig.~\ref{fig:8}a) of their first neighbors, so we have a discrete translation symmetry on space ($\hat x$) and a continuous one on time (Fig.~\ref{fig:2}a). The corresponding fundamental mode is $M_s(k_s,0)$.

\item[(ii)] The transition from ST to TW breaks the parity symmetry ($x \rightarrow -x$) and the temporal continuous translation symmetry. In the TW regime (Fig.~\ref{fig:3}b), the new discrete translation symmetry on space is given by a wavelength $\lambda_{TW} \approx 3/2 \lambda_s \equiv \lambda_{3s/2}$ (i.e. at $d=$ 4mm see Fig.~\ref{fig:8}a). This result differs from the doubling period obtained in a previous work~\cite{Burguete03} because the precision for both, the experimental techniques and the analytic data processing, has been remarkably improved. The hotspots decide to travel at the same time towards a privileged direction selected by the system at a frequency $\omega_{TW}$, which varies linearly with $\Delta T_v$ (Fig.~\ref{fig:8}b). The mean period is approximately 25 s. The velocity at which hotspots travel is the phase velocity $v_{\phi}=\lambda_{3s/2}\;\omega_{TW}/2\pi$, because we have measured that the linear group velocity verifies $v_g =\mathrm{d}\omega/\mathrm{d}k \approx$ 0, as it will be shown further on.
So the phase velocity increases linearly with an increasing control parameter $\Delta T_v$ as we go beyond the threshold of the secondary bifurcation to TW. 
The corresponding fundamental mode is (either right or left): $M_{v\pm}(k_{2s/3},\pm\omega_{TW})$.
\item[(iii)] In the ALT pattern (Fig.~\ref{fig:3}c) the hotspots are oscillating in counterphase with a new discrete spatial translation symmetry given by $\lambda_{ALT} \approx 2 \lambda_s =\lambda_{2s} $ (i.e. at $d=$ 4mm see Fig.~\ref{fig:8}a) in agreement with previous experimental work~\cite{Burguete03}. So hotspots behave as oscillators that by pairs interchange their positions to recover their initial positions after a cycle of frequency $\omega_{ALT}$. This frequency is almost constant as we increase $\Delta T_v$ (Fig.~\ref{fig:8}b). The mean period is about 23 s. So, for an ascending sequence the phase velocity keeps constant as we go beyond the threshold of the secondary bifurcation to ALT. This pattern is strongly nonlinear as the result of the competition between the fundamental modes of a resonant triad: $M_s(k_s,0)$ and $M_{v\pm}(k_{s/2},\pm\omega_{ALT})$. The stationary mode $M_s(k_s,0)$ characteristic of the cellular pattern ST has been a sort of ``{\it restored}''.
\end{itemize}

In this array of coupled oscillators there may exist localized discontinuities that correspond to boundaries. These boundaries or fronts break the dynamics into two different groups of oscillators, each one of them with a great spatiotemporal coherence. On the spatiotemporal diagram $S(x,t)$ these regions are called localized domains. Such is the case of the existence of one or various localized domains in the ALT pattern coexisting with the ST pattern as a ``mixed'' state (Fig.~\ref{fig:3}d). This domains usually collapse, they have fluctuating boundaries with variable spatiotemporal areas. This pattern from now on is designated as ST/ALT.\par

\subsection{Primary and secondary instabilities}
 
\begin{figure*}
\includegraphics[width=10cm]{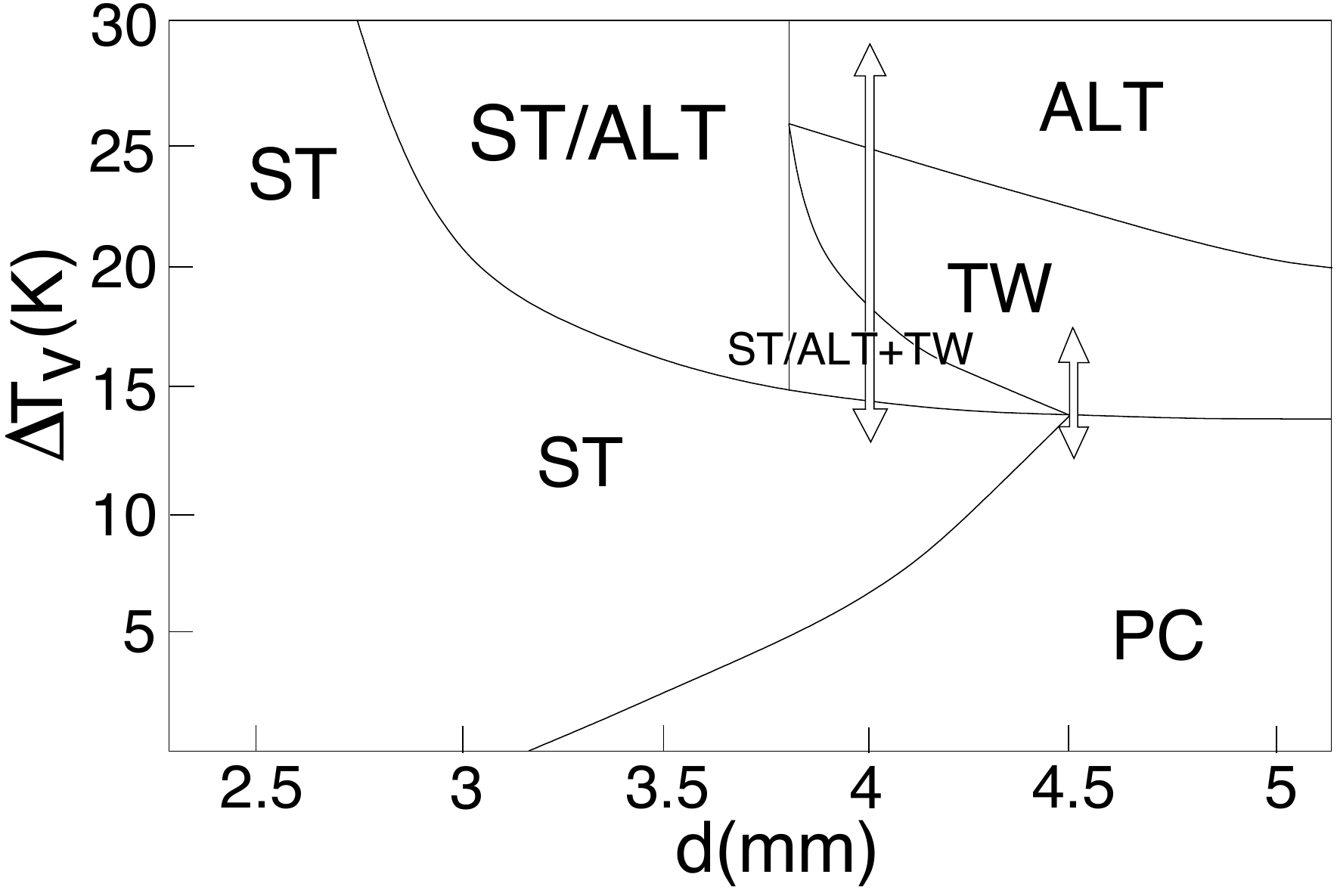}
\caption{\label{fig:4} Stability diagram ($d$, $\Delta T_v$) built from the spatiotemporal diagrams $S(x,t)$. PC and ST are stationary patterns, TW and ALT are oscillatory patterns and ST/ALT is the mixed pattern. ST/ALT + TW defines the subcritical region where a stationary front connects ST/ALT and TW. Double arrows correspond to the stepped sequences of measurements reported.}
\end{figure*}

A smooth increase of the control parameter $\Delta T_v$ can lead from the homogeneous pattern PC to the onset of two different kinds of {\it primary bifurcations} depending on the depth of the layer $d$ (see the stability diagram in Fig:~\ref{fig:4}):
\begin{itemize}
\item[(i)] For $d\le$ 3.5 mm, as soon as $\Delta T_v \not=0$ the ST pattern becomes unstable. For this pattern we find a higher threshold as we increase the depth $d$ from 3.5 mm to 4.5 mm.

\item[(ii)] For $d\ge$ 4.5 mm, the TW pattern becomes unstable from the primary roll PC. This bifurcation is thought to be supercritical concerning the experimental fact that the TW pattern appears simultaneously along the cell from an homogeneous pattern at d = 5 mm. 
\end{itemize}

For the explored depth range $d=$ 2.5-5 mm, there are the following pathways to {\it secondary bifurcations} (see the stability diagram in Fig:~\ref{fig:4}):
\begin{itemize}
\item[(i)] {\bf Thick layers}: For 4.5 $\le d \le$ 5 mm and further from the threshold of the primary bifurcation to TW, the system undergoes a new bifurcation to the ALT pattern. The relative amplitudes verify $|A_{v-}| \approx |A_{v+}|$ and $|A_{s}| \approx |A_{v\pm}|/2$. 

\item[(ii)] {\bf Intermediate layers}: For 4 $\le d \le$ 4.5 mm, the system becomes unstable from ST to TW via a bistable regime. This bistable regime consists of a mixed pattern, ST/ALT, coexisting with the new TW pattern. The front connecting these two regimes (ST/ALT and TW) is stationary for the front velocity is $v_p\approx$ 0.01 mm/s. This front velocity is determined by measuring the slope of the front defined by the new mode, which becomes unstable, at the spatiotemporal diagram. In order to study this bistability we focus on this second instability to TW for two intermediate depths: 4 and 4.5 mm. Beyond the TW pattern we report a new secondary bifurcation towards the ALT pattern at $d =$ 4 mm.

\item[(iii)]{\bf  Thin layers}: For $d \le$ 3 mm, from the basic flow pattern ST, a cascade of secondary bifurcations takes places towards a pattern of localized domains over the basic pattern ST. These results will be reported elsewhere. The first instability from the basic pattern ST is the mixed ST/ALT pattern (Fig.~\ref{fig:3}d). The quantitative analysis shows that $|A_{v\pm}|$ keeps constant inside these domains along a sequence for increasing steps.
\end{itemize}

For thinner layers ($d \le$ 2.5 mm) and high temperatures ($\Delta T_v \ge$ 55 K) the heating looses its linear localization and hence it turns into an homogeneous 2D heating. Under these conditions, the system gives birth to an hexagonal pattern arranged symmetrically with respect to HL, similar to the classical B\'enard-Marangoni convection.\par
\begin{figure}
\includegraphics[width=5cm]{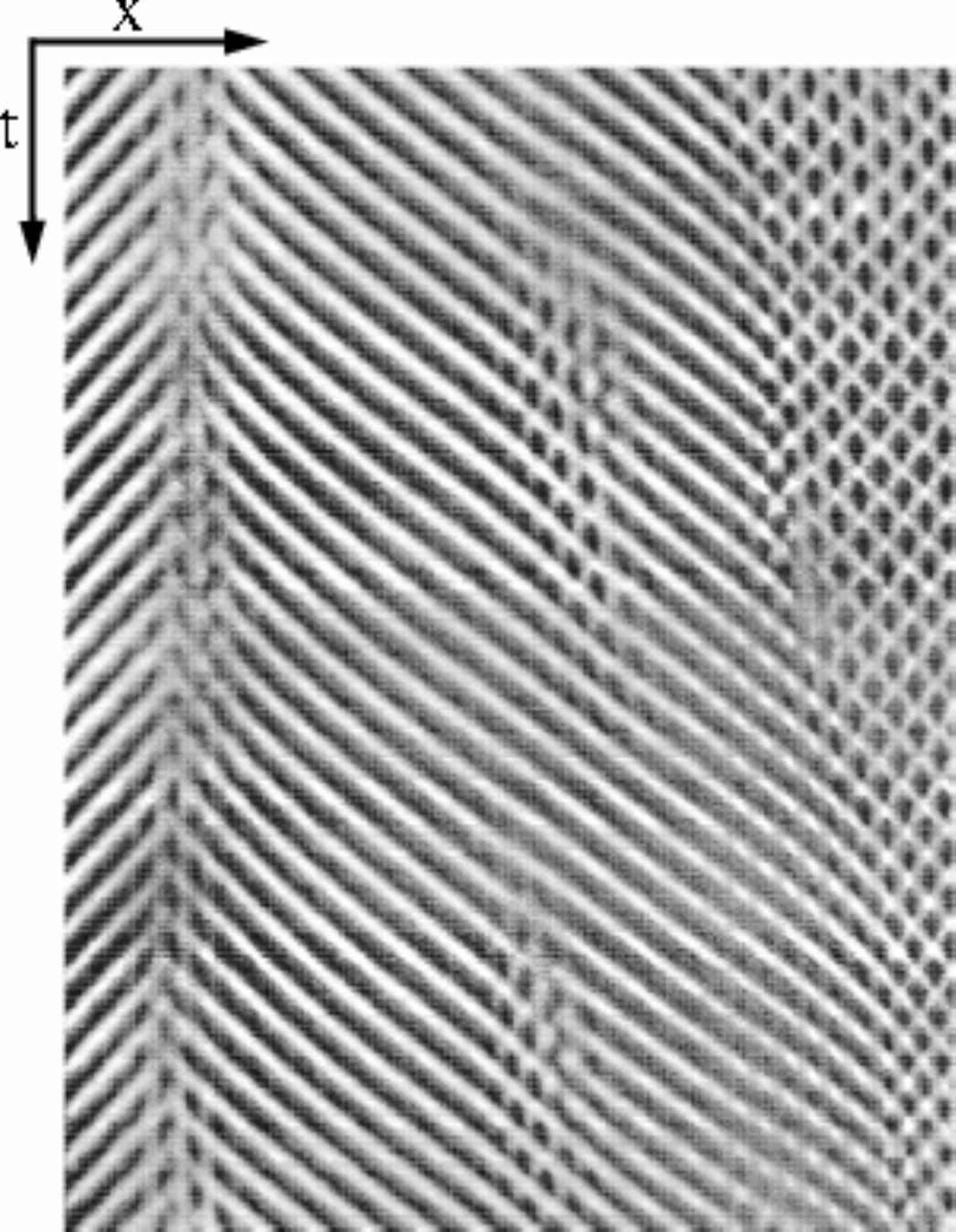}
\caption{\label{fig:5} Spatiotemporal diagram of TW ($d =$ 4 mm, $\Delta T_v=$ 20 K) with localized drifting domains in the ALT pattern in a transient (for an upward jump with step  $\Delta T_h=$ 4 K) and the presence of a source of TW in the left hand side.}
\end{figure}

  
Beyond the threshold of the last secondary bifurcation, for each $d$ and until the maximum value of $\Delta T_v$ applied, the dynamics of the system still keeps the characteristic patterns, but with the presence of defects of the type of dislocations with topological charge +1. For transient regimes we have observed the existence of pulses, sources, sinks and localized drifting domains (see Fig.~\ref{fig:5}). Also, in transients we can find counterphase oscillating patterns (optical modes). For steady patterns when there is a domain boundary (or front) connecting the patterns TW and ST/ALT, these oscillating patterns are present as well. As it is shown later on, oscillatory patterns are similar to the ALT pattern on account of the fundamental modes but with different amplitude rates.\par

From now on, our aim is to study the subcritical nature of the system in the secondary bifurcation towards TW (with a backward study research looking for an hysteresis cycle) and towards ALT.

\subsection{Secondary bifurcation to traveling waves and further}

\begin{figure*}
\includegraphics[width=16cm]{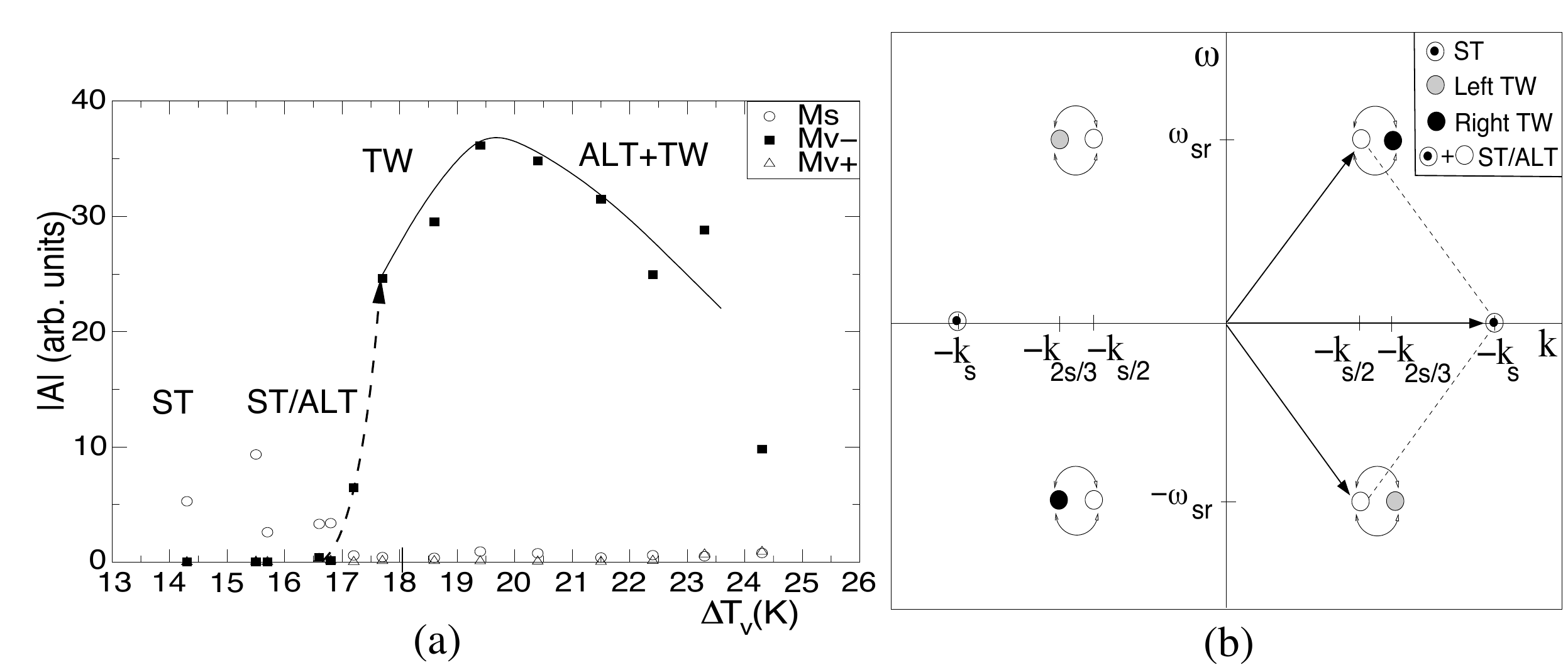}
\caption{\label{fig:7}(a) Bifurcation diagram at $d=$ 4 mm for ascending steps of 0.5 K. The continuous line is a guide to the eye. The discontinuous arrow show the upward jump of the amplitude mode $M_{v-}$ at $\Delta T_v = 17.7$ K ($\varepsilon =$ 0) for the secondary bifurcation to TW. Far from this threshold at $\Delta T_v \approx$ 19.2 K the amplitude of the mode $M_{v-}$ decays in favor of the counterpropagative mode $M_{v+}$ representing the advance of the front which splits the patterns TW and ALT. (b) Sketch of the mechanism of activation of resonant triad in the ALT domains (mixed ST/ALT pattern) by the emerging TW modes. $\omega_{sr}$ is the frequency at the subcritical interval ($-\varepsilon_c \le \varepsilon \le 0$).}
\end{figure*}

For intermediate layers, when the system approaches the threshold from the cellular pattern ST to the secondary bifurcation to TW we define the reduced control parameter as $\varepsilon = \Delta T_v/ \Delta T_{vc}^a - 1$. $\Delta T_{vc}^a$ is defined as the critical parameter at which the system becomes unstable towards the new state, TW, for an ascending sequence of measurements with the minimum step $|\Delta T_h|$. The amplitude of the corresponding unstable mode $|A_{v-}|$ grows abruptly at this point meanwhile the amplitudes $|A_s|$ and $|A_{v+}|$ decay to zero (i.e. see Fig.~\ref{fig:7}a). The subcritical parameter for the hysteresis cycle is given by $\varepsilon_c = \Delta T_{vc}^d/ \Delta T_{vc}^a - 1<0$, where $\Delta T_{vc}^d$ is the critical parameter (corresponding to a descending sequence) at which the system returns to the original state, ST. Thus, the range $\varepsilon_c \le \varepsilon \le 0$ is the subcritical interval. At $d=$ 4 mm, for an ascending sequence with step 0.5 K from the amplitudes diagram (Fig.~\ref{fig:7}a) we have measured $\Delta T_{vc}^a =$ 17.7 K. To verify the existence of hysteresis we have worked with ascending and descending steps of 0.3 K in the subcritical interval and we have obtained $\Delta T_{vc}^a =$ 16.4 K and $\varepsilon_c =$ -0.07.\par
Close to the codimension-2 point (at $d=$ 4.5 mm) describing a cycle with step 0.3 K and with the same processing protocol we have determined $\varepsilon_c =$ -0.02.

\begin{figure*}
\includegraphics[width=16cm]{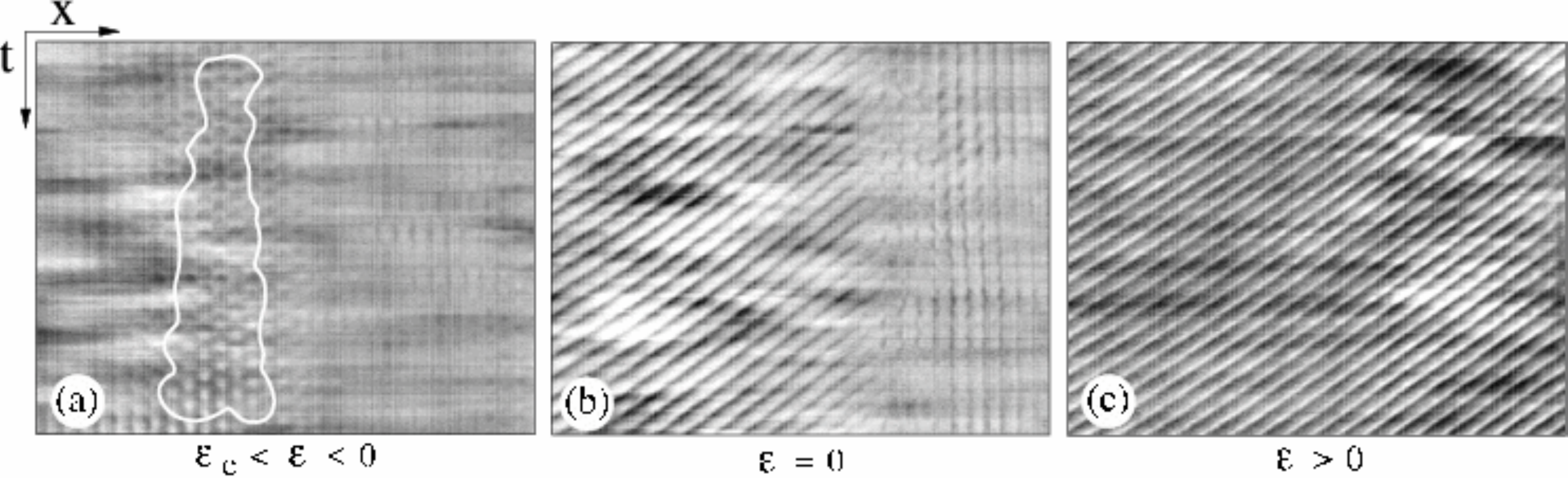}
\caption{\label{fig:6} Spatiotemporal diagrams corresponding to the secondary bifurcation to TW at $d=$ 4 mm (for an ascending sequence): (a) the mixed ST/ALT pattern belongs to the subcritical interval, the boundary of the ALT domain has been highlighted (filtering with $\mu=$ 0.5); (b) the bistable pattern: ST/ALT and TW at the threshold, these patterns are connected by a stationary front; (c) the TW pattern above the threshold.}
\end{figure*}

In a correlative ascending sequence for the secondary bifurcation to TW (Fig.~\ref{fig:6}):
\begin{itemize}
\item[-] For $\varepsilon_c \le \varepsilon \le 0$ the mixed pattern ST/ALT is present. In the localized domains of the ALT pattern the competition between the traveling modes $M_{v+}(k_s/2,\omega_{ST/ALT})$ and $M_{v-}(k_s/2,-\omega_{ST/ALT})$, with similar amplitudes, produce a resonant nonlinear interaction with the stationary mode, $M_s(k_s,0)$.  
\item[-] For $\varepsilon =$ 0 the system bifurcates towards TW at the same time that coexists with the original ST/ALT pattern, this is the bistable regime. In this regime a discontinuity of the amplitude of the traveling mode $M_{v-}(k_{2s/3},-\omega_{TW})$ has been produced in the corresponding domain, the TW pattern. The front velocity is approximately zero, $v_p \approx$ 0.01 mm/s.
\item[-] For $\varepsilon >$ 0 the whole system has bifurcated to TW without the existence of boundaries, neither sources nor sinks, so $M_{v-}(k_{2s/3},-\omega_{TW})$ is a global unstable mode.
\end{itemize}

\begin{figure*}
\includegraphics[width=16cm]{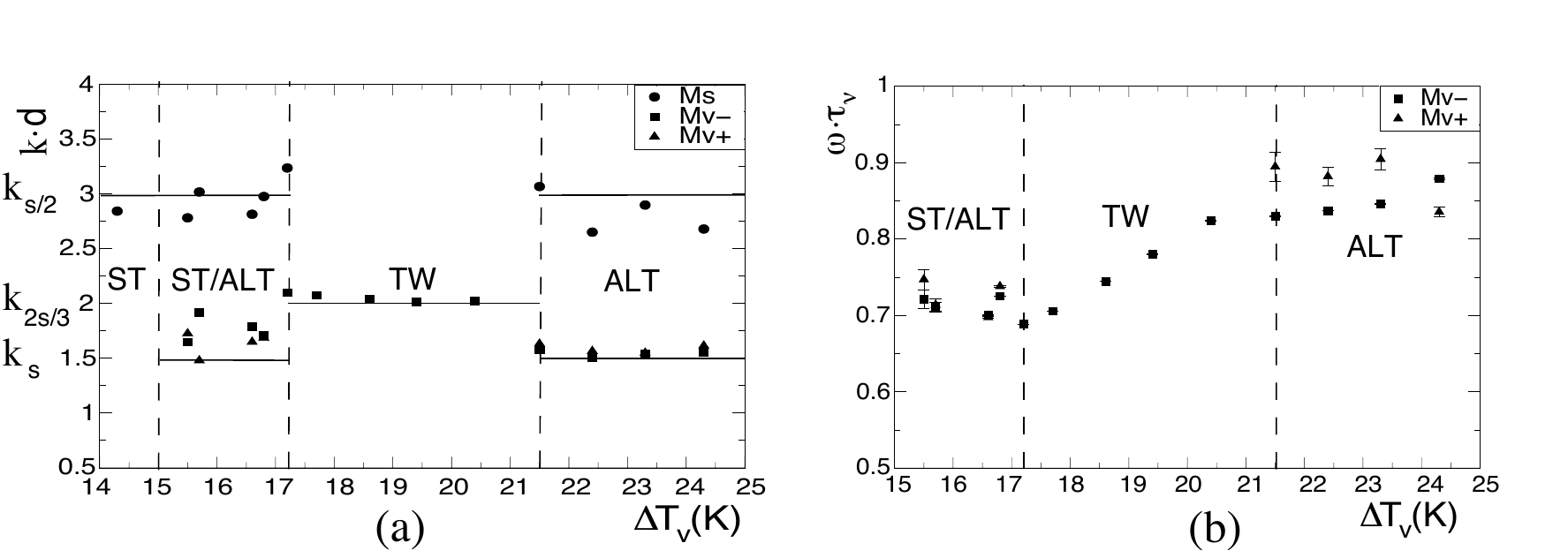}
\caption{\label{fig:8}At $d =$ 4 mm evolution $vs.$ $\Delta T_v$ (with step $\Delta T_h=$ 0.5 K) of: (a) the dimensionless wavenumbers (horizontal continuous lines correspond to the averaged wavenumbers for the same patterns at 3 mm and 4 mm) and (b) the frequencies ($\tau_\nu=$ 3.2 s) for each involved fundamental mode: $M_s$(stationary mode), $M_{v\pm}$ (right-left traveling modes). Vertical dashed lines separate different patterns.}
\end{figure*}

\begin{figure*}
\includegraphics[width=13cm]{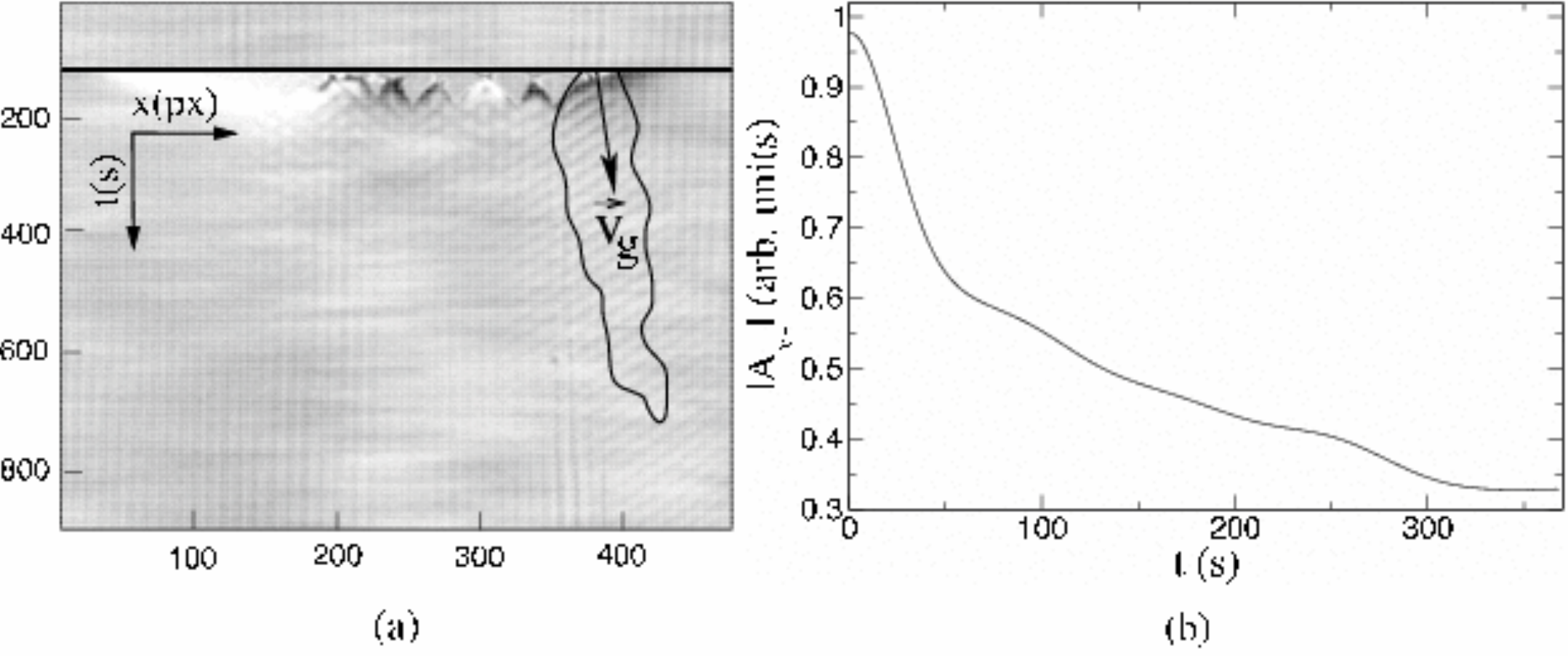}
\caption{\label{fig:9}(a) Spatiotemporal diagram with a localized mechanical perturbation close to the threshold to TW at ($d=$ 4 mm, $\Delta T_v =$ 15.5 K). The boundary of the transient TW domain (filtered with $\mu =$ 0.37) has been sketched; (b) Attenuation of the module of the traveling mode $M_{v-}$ inside the transient domain.}
\end{figure*}

For this secondary bifurcation to TW if we artificially excite (perturbing locally the free surface) the basic state ST just below $-\varepsilon_c$ at ($d=$ 4 mm, $\Delta T_v =$ 15.5 K), we observe the birth of a TW domain (see Fig.~\ref{fig:9}a). This domain vanishes after a while because of the attenuation of the amplitude of the unstable traveling mode (Fig.~\ref{fig:9}b) with a front velocity of $v_p \approx$ 0.17 mm/s. This is what is expected for a negative growthrate when we are below and close to the threshold. For these control parameter values  the TW pattern is said to be convectively unstable with a negative group velocity: $|v_g|\approx$ 0.12 mm/s, and then, the phase velocity verifies $v_\phi\approx$ -3.5 $v_g$. If we perturb the system above and close to the threshold, the group velocity is null.\par 

\begin{figure*}
\includegraphics[width=16cm]{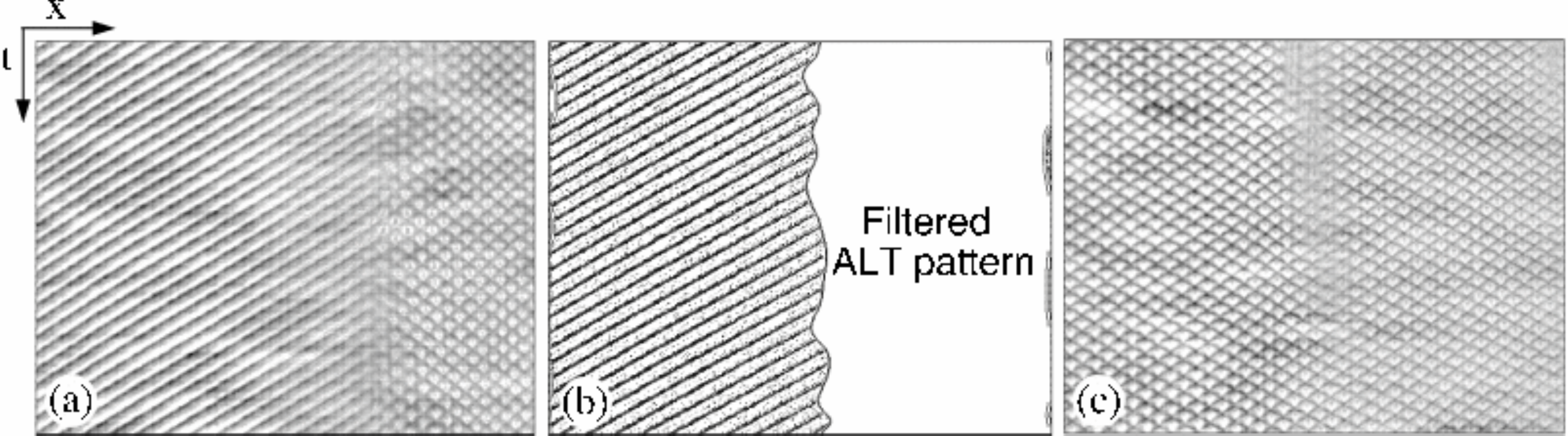}
\caption{\label{fig:10}Spatiotemporal diagrams corresponding to a secondary bifurcation to ALT above the threshold: (a) There is a front connecting the TW and ALT patterns at $\varepsilon^\star=$ 0.11; (b) Contrasted profile of the front (The left traveling mode has been filtered with $\mu =$ 0.37) at $\varepsilon^\star=$ 0.15; (c) the ALT pattern has invaded the whole cell far from the threshold at $\varepsilon^\star=$ 0.38.}
\end{figure*}

Further from threshold at $d =$ 4 mm the basic state of TW bifurcates to the new state, ALT, via a dynamics of stationary fronts at the asymptotic states (see the spatiotemporal and filtered diagrams in Fig.~\ref{fig:10}a,b). For these fronts we have measured a front velocity of $v_p \approx$ 0.01 mm/s. We define a new threshold for an ascending sequence (with step 1K) as $\varepsilon^\star = \Delta T_v/\Delta T_{vc} - 1= 0$ where at $\Delta T_{vc}$ the ALT pattern appears for the first time ($\Delta T_{v}=21.5$ K). From this point on and as far as we increase $\Delta T_v$, the front invades the TW pattern until at $\varepsilon^\star=$ 0.29 the whole system dynamics has transitioned to the ALT state (Fig.~\ref{fig:10}c). In this transition we have measured a finite jump for the amplitudes of the traveling modes at the threshold: $|A_{v\pm}|(ALT)\approx 3|A_{v-}|(TW)$ while the amplitude of the stationary mode fulfils: $|A_{s}|(ALT)\approx |A_{v\pm}|/2(ALT)$.\par
\begin{figure}
\includegraphics[width=8cm]{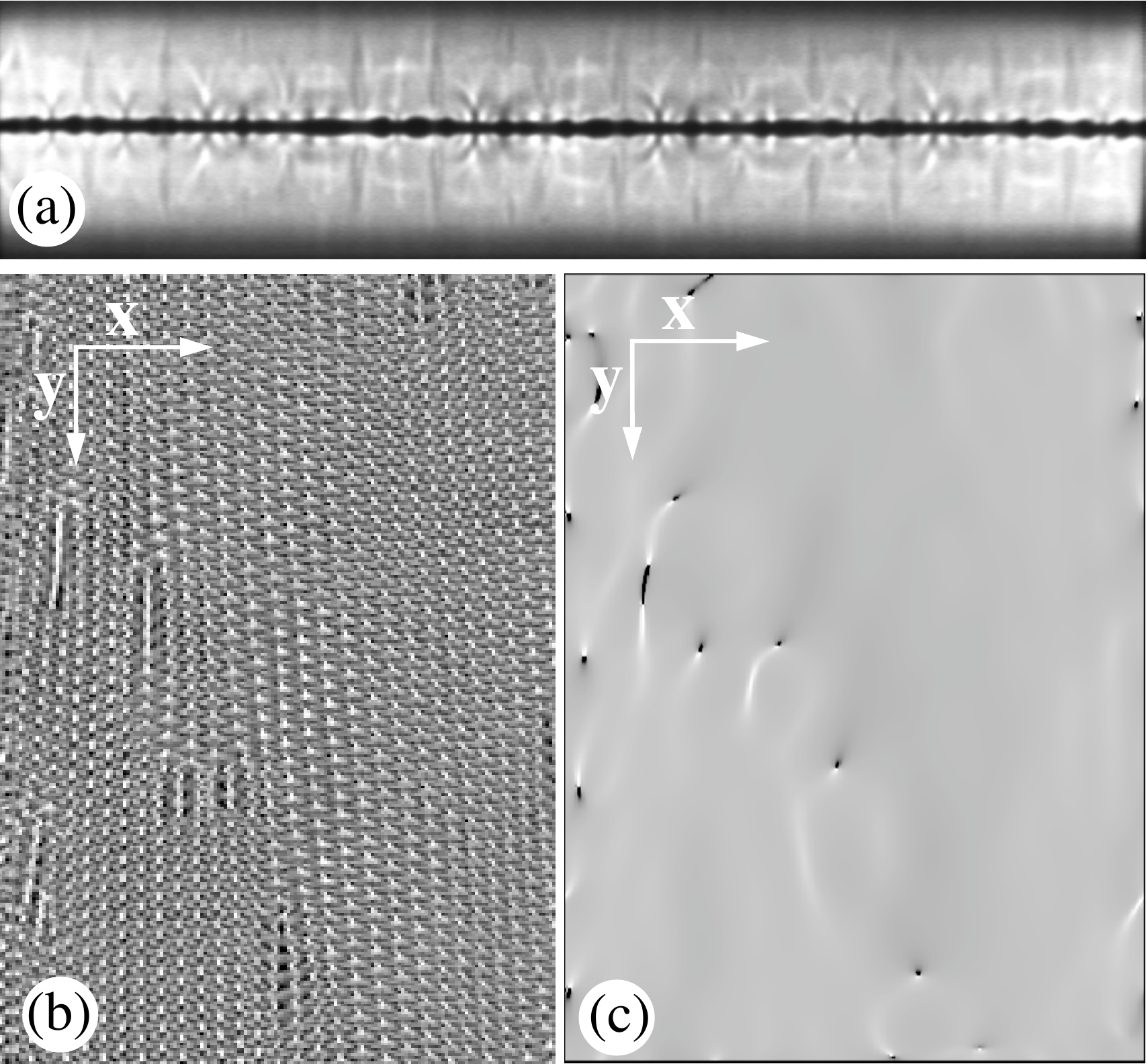}
\caption{\label{fig:11}ALT pattern with the presence of defects at $d =$ 5 mm and $\Delta T_v= 62$ K: (a) The shadowgraphy image on the screen; (b) the spatiotemporal diagram; (c) phase gradient of the stationary mode $M_s$, the discontinuities correspond to dislocations.}
\end{figure}

For higher $|\Delta T_v|$, the turbulent ALT pattern has a dynamics of defects (dislocations) (Fig.~\ref{fig:11}a,b). Buoyancy plumes can be observed in the shadowgraphy image, they are detached along the perpendicular axis to HL. On the phase gradient in Fig.~\ref{fig:11}c we can observe how these defects are spontaneously generated, adjusting locally the wavenumber of the pattern.\par

\begin{figure*}
\includegraphics[width=16cm]{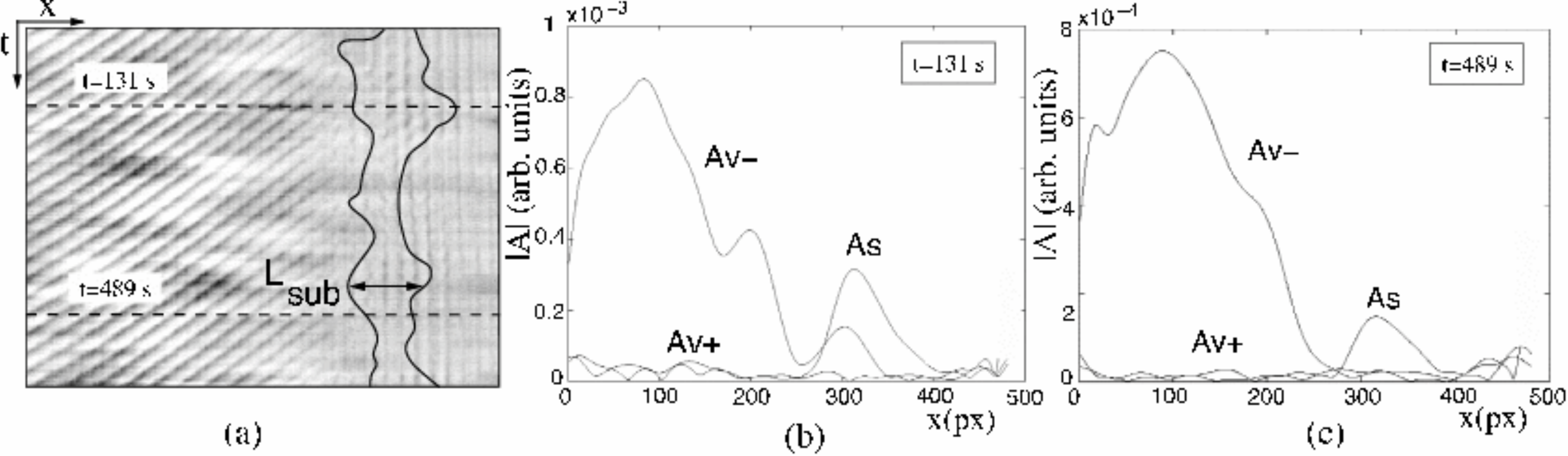}
\caption{\label{fig:12}(a) For ($d=$ 4 mm, $\Delta T_v= 16$ K) the spatiotemporal diagram of a bistable pattern in the secondary bifurcation to TW where the previous pattern ST/ALT and TW coexist. The boundary domain where the stationary mode $M_s$ is coupled with the traveling modes $M_{v\pm}$ has been sketched (with $\mu=$ 0.37). $L_{sub}$ is the average width of the boundary domain. The evolution of the amplitudes of the fundamental modes for the same spatiotemporal diagram (b) at $t=131$ s and (c) $t=489$ s. The growth of $A_s$ corresponds to the resonant interaction at the front.}
\end{figure*}

Because of the subcritical behavior of the system in the secondary bifurcation to TW, nonlinear interactions between the original (ST/ALT) and the new (TW) patterns in the bistable regime (Fig.~\ref{fig:12}a) are expected. Indeed the boundary is likely to show this nonlinear behavior by resonant interactions between the competing modes. In consequence, we focus not on the evolution of each of the coexisting domains, but on the overlapping between the original mode $M_s$ (representing the mixed ST/ALT pattern) and the new mode $M_{v-}$ (representing the new pattern TW) at the front. For this analysis the signal $S(x,t)$ has been filtered with $\mu=$ 0.37. The average width of the front is defined as a  subcritical length, $L_{sub}$, where the mode $M_s$ is interacting nonlinearly with the others ($M_{v\pm}$). \par
To explain the resonance of the amplitude $|A_s|$ as a nonlinear coupling between the triad ($M_s$, $M_{v-}$ and $M_{v+}$) along $L_{sub}$, we study the profile of the modulus of the amplitudes of the fundamental modes in the direction perpendicular to the front, so we can check the amplitude gain of $|A_s|$ at $t=$ 131 s (see Fig.~\ref{fig:12}b) and the minimum value of $|A_s|$ at $t=$ 489 s (see Fig.~\ref{fig:12}c) . The mode $M_s$ is coupled with the propagative modes in the boundary where the unstable mode $M_{v-}$ penetrates beyond the stationary front, and therefore in this region the modulus $|A_{s}|$ will be greater. Filtering with $\mu = 0.37$ and for $t =$ 131 s, in the region $L_{sub}$, we obtain the following results $|A_s|/|A_{v-}| \approx$ 2 and $|A_{v-}|/|A_{v+}| \approx$ 6. In the boundary the kinematic resonant conditions~\cite{Hammack93} for the wave numbers: $0.7774 \mbox{ mm}^{-1}\approx 0.4438 \mbox{ mm}^{-1} + 0.3135\mbox{ mm}^{-1}$; and for the frequencies: $0.0030\mbox{ s}^{-1}\approx - 0.2102\mbox{ s}^{-1} + 0.2210 \mbox{ s}^{-1}$ are fulfilled.

For each one of these data the corresponding error from the analytic processing is negligible considering the experimental error due to temperature inhomogeneities.\par
The corresponding oscillatory pattern that shapes the boundary contour is very similar to the previously mentioned for the optical modes. This pattern differs from the ALT uniquely on account of the different ratios between amplitudes.\par
We can interpret the growth of the modulus $|A_s|$ (Fig.~\ref{fig:12}b) as a positive energy transference from the propagative mode $M_{v-}$ (TW) to the stationary one $M_{s}$ (ST/ALT). At $t =$ 489 s the amplitude of $M_s$ has been attenuated regarding the previous result at $t=$ 131 s, meanwhile for the propagative modes the ratio of amplitudes accomplishes: $|A_{v-}|/|A_{v+}| \approx$ 1 (see Fig.~\ref{fig:12}c). \par

Concerning the correlation length for the subcritical instability to TW, we may associate the subcritical length $L_{sub}$ with an attenuation length of the new pattern (TW) inside the original one (ST/ALT). Nevertheless, the evolution of the averaged front width, as a correlation length parameter, has not revealed any evidence at the onset of this transition. Fluctuations of the amplitude $|A_s|$ are caused by the nonlinear resonance at the front, so this fact has to be considered to implement the most suitable strategy for the data analysis in bistable regimes. \par

\section{DISCUSSION AND CONCLUSIONS}
\label{sec:discussion}

\subsection{An extended 1D convective system}
 
The two counterotating rolls (PC) in this experiment remain for any $\Delta T_v$, so we might compare the basic cellular pattern ST (which is a transverse cellular pattern regarding the wave vector of the primary roll) with other systems with similar behavior such as the lateral heating experiment in a rectangular cell with $Pr=$ 10~\cite{Burguete01}. In this experiment, the fluid dynamics for deep layers corresponds to stationary rolls with a wavevector perpendicular to the temperature gradient, this pattern belongs to the type of stationary rolls predicted by Smith \& Davis~\cite{Smith83}. These stationary rolls become unstable over the underlying ``unique'' roll which invades the whole cell. For larger horizontal temperature gradients the system becomes unstable via an oscillatory pattern of optical modes. This pattern has a strong affinity with the resonant triad of the ALT regime where the counterpropating modes $M_{v \pm}$ have restored the stationary mode $M_s$. \par

For each pattern, the dynamics of the central 31 cm of the cell is equivalent to the dynamics of an array of 56 oscillators (from a total of 75) interacting with their neighbors. We are studying the collective behavior of a set of macroscopic particles (hotspots), a 1D array of nonlinear coupled oscillators that become unstable. Experimentally we are performing ``coarse-graining'' when we associate to the hotspots on HL the microscopic variables of the convective system like the thermal fluctuations. Besides, as the physical aspect ratios for the cellular pattern at $d =$ 3 and 4 mm are respectively $\Gamma_\lambda = L_x/ \lambda_s =$ 75 and 59, we may consider our system to be extended. \par

According to the experimental results the system may suffer two different bifurcations to TW: subcritical and supercritical, depending on the basic state and thus, on $d$. In consequence, from the homogeneous state PC the system undergoes an oscillatory supercritical Hopf bifurcation that can be theoretically described by the cubic complex Ginzburg-Landau equation. Experimentally, this kind of supercritical bifurcation has been found in many other extended systems~\cite{Burguete01,Pastur03} and references therein. Meanwhile the system is behaving subcritically for $d \le$ 4.5 mm. 
 
From these results we may infer that as the dynamic Bond number is $Bo_D \propto d^2$, when $Bo_D$ decreases (as we move to the left in the stability diagram of Fig.~\ref{fig:4}) a subcritical nature of the secondary bifurcations arises. So in this experiment subcriticality may be linked to the fact that thermocapillary effects are becoming more important the thinner the layer is. Subcriticality via a Hopf bifurcation to TW has been reported experimentally for example in binary mixture of fluids~\cite{LaPorta96}.\par
 
Our results for the secondary transition to TW show that $\lambda_{TW}/\lambda_s \approx 3/2$. Under the light of this result we leave an open question to derive a suitable model as the one developed by L.Gil~\cite{Gil00} for a 1D system. This model, for a Floquet exponent $q=1/2$, differs from CI because the constraint over equal amplitudes of the right-left envelope functions (propagating modes) is relaxed. Hence this model can describe mixed patterns and localized coherent structures from a secondary instability, fortunately not too close to the threshold because a phase shift mode has not been disregarded~\cite{Gil99}. In consequence, a model like this could be tried out for a Floquet exponent $q=2/3$. 
Meanwhile, in the secondary bifurcation to ALT we show that $\lambda_{ALT}/\lambda_s \approx 2$, for this reason this instability belongs to the spatial period-doubling oscillatory bifurcation predicted by the CI theory, although it is build in the framework of a weakly nonlinear regime.

\subsection{Stability diagram}

Experimental studies for silicone oils with $Pr \ge \mathcal O(10)$~\cite{Krishnamurti73} show that each regime in the stability diagram is well defined because it is confined in large regions of the parameter space. In other words, there is a long pathway towards turbulence with a great diversity of regimes. Therefore consecutive instabilities can be evaluated separately. Our system (Pr$\approx$ 75) exhibits about eight different regimes in the explored space of the control parameters ($d$, $\Delta T_v$) including bistable regimes.\par

For $d <$ 4 mm the quantitative study of the evolution of the amplitudes ($|A_s|$ and $|A_{v\pm}|$) does not allow to discover the nature of the transition from the ST pattern to the mixed ST/ALT pattern . We may assert that for $d<$ 4 mm the basic pattern ST never looses its stability. On the other hand, we have measured that the surface occupied by the localized domains in the ALT pattern with respect to the total surface of the spatiotemporal diagram is variable as we increase $\Delta T_v$.\par 

The system is bistable in the range of intermediate layers (4 mm $\le d \le$ 4.5 mm), where a ST/ALT domain may coexist with a TW domain. Bistability is the reason why, in the secondary bifurcation to TW, two states with high spatiotemporal coherence are solution of the system for the same range of the control parameters $(d, \Delta T_v)$. In Fig.~\ref{fig:7}b it has been sketched how the mixed ST/ALT pattern can be considered as a ``transitional'' pattern selected by the system for the subcritical bifurcation to TW. As we approach the bistable regime and until the system has fully transitioned to TW, the emerging traveling modes with wavenumbers $k_{2s/3}$ play a destabilizing role towards the nearby oscillating modes with wavenumbers $k_{s/2}$. As the frequencies in the subcritical interval are roughly the same for both oscillatory modes (see Fig.~\ref{fig:8}b), $\omega_{sr}$, close to the subcritical parameter $-\varepsilon_c$ the stationary mode, $M_s(k_s,0)$, activates the resonant ALT modes, $M_{\pm v}(k_{s/2},\omega_{sr})$, before the TW modes, $M_{\pm v}(k_{2s/3},\omega_{sr})$, which are responsible for the next global bifurcation. Attending to the first nonlinear resonant criterion, a resonant interaction cannot be triggered between the stationary and the traveling modes. For the propagative modes in this mixed ST/ALT pattern, we have measured (Fig.~\ref{fig:8}a) a finite jump of wavelengths ($\delta k\cdot d \approx$ 0.25) in comparison with the average wavelength of the propagative modes in the ALT pattern, this difference is smaller than the experimental error.\par
 
As we get closer to the codimension-2 point the subcritical parameter $\varepsilon_c$ tends to zero. At this critical point and depending on the way we approach it, the system may bifurcate to TW from two different patterns: homogeneous (PC) or cellular (ST). From $d\approx$ 4.5 mm the threshold for the PC towards the ST pattern gradually decreases until $d\approx$ 3, where the ST pattern settles with no threshold. This experimental result is in agreement with the theoretical prediction (a linear stability analysis) for a similar experimental setup~\cite{Mancho97} that shows that, as far as the Gaussian profile of the temperature field flattens a smaller threshold for the ST pattern is expected. \par

From the PC pattern and for $d \ge $ 4.5 mm, the system undergoes a supercritical bifurcation towards TW. At the threshold, we may think of a whole array of hotspots that cooperatively decide to travel towards a privileged direction, so the correlation length is supposed to diverge and consequently, for this supercritical transition, there are no fronts (even though the presence of fronts is not necessarily linked to subcriticality).\par

From the results obtained for the secondary bifurcation to ALT, we can assert that bistability is the cause for the coexistence of the patterns TW and ALT. Furthermore, it is the cause for the existence of localised drifting domains in the ALT regime in transients. This is the case for an abrupt upward jump (in terms of $\Delta T_v$) above the TW threshold where the presence of drifting domains in ALT have locally broken the symmetry imposed by the basic TW pattern (Fig.~\ref{fig:5}).\par

It should be pointed out that the ALT pattern is strongly nonlinear and far from the threshold it shows a typical turbulent dynamics of defects (see Fig.~\ref{fig:11}). In the threshold to the ALT pattern and above, the system always selects the two counterpropagative modes $M_{v\pm}(k_{s/2},\omega_{ALT})$ which restore the stationary one $M_{s}(k_s,0)$ by nonlinear resonance. By contrast, at the threshold of the secondary bifurcation to TW there are two possible modes that may become unstable, $M_{v-}$ or $M_{v+}$, only one of them will be selected by the system due to nonlinear effects. \par

In both secondary bifurcations the fronts that connect bistable patterns are stationary. Certainly, for intermediate layers, the predominant thermograviational effect together with the underlying cellular pattern (for ST/ALT and ALT) are the major causes for the strong locking that restrains uniformly the diffusion of a new pattern into the original one at the asymptotic states. On the contrary, for thinner layers ($d<$ 4 mm) the thermocapillary effects probably account for the fluctuating fronts in the ST/ALT pattern. \par

 \subsection{Convective/absolute instability towards traveling waves and the subsequent alternating pattern}
 
For an experimental 1D convective system which undergoes a secondary bifurcation to TW, we have shown how the existence of hysteresis can be tested studying the amplitude of the fundamental traveling modes ($|A_{v+}|$ or $|A_{v-}|$ ). According to the existence of a subcritical interval for the secondary bifurcation to TW, at $d=$ 4 mm the finite jump of wavenumbers and frequencies is produced at $\Delta T_{vc}^a =$ 17.2 K (Fig.~\ref{fig:8}a,b), before the jump of the amplitude of the mode $M_{v-}$ at $\Delta T_{vc}^a =$ 17.7 K (Fig.~\ref{fig:7}a) for an ascending sequence (with step 0.5 K). Close to the threshold these results might not be surprising, if we follow the usual form \cite{Coullet89}, under the symmetries of the problem (parity-breaking bifurcation), for the coupling between the slowly varying phase $\phi$ and the amplitude $\mathcal A$ of the traveling mode:

\begin{eqnarray}
\mathcal A_t=\mathcal A_{xx}+f\left(\mathcal A\right)+\alpha \phi_x \mathcal A+ \cdots\;,
\label{eq:one} 
\\
\phi_t=\phi_{xx}+\beta \mathcal A+ \cdots\;,
\label{eq:two} 
\end{eqnarray}

where $\alpha$ and $\beta$ are the coupling parameters. In our subcritical bifurcation to TW, $f(\mathcal A)$ is expressed in terms of a fifth-degree polynomial of $\mathcal A$. 
In accordance to equation Eq.~(\ref{eq:one}), when we mechanically perturb the fluid the excitation of the amplitude of the traveling mode (Fig.~\ref{fig:9}) implies a coupling between the critical mode associated to the secondary bifurcation ($M_{v+}$ or $M_{v-}$) with the phase gradient of the basic state ST, that is: $\mathcal A_t= \alpha \phi_x \mathcal A+ \cdots\;$. The attenuation of the traveling mode in Fig.~\ref{fig:9}b shows that the domain in TW disappears because of a relaxation dynamics for the unstable mode.\par

We have observed the bistable pattern (coexistence of the ST/ALT and TW patterns) for positive and small values of $\varepsilon$ because our measurements are restricted by the size of the step $\Delta T_h$. From the experimental results at the threshold, if we compare two ascending sequences with different steps 0.5 K (Fig.~\ref{fig:7}) and 0.3 K, we find that the discontinuous jump in the amplitude $|A_{v-}|$ is produced before with the smaller step at $\Delta T_{vc}^a =$ 16.3 K ($\Delta T_{vc}^a =$ 17.7 K with step 0.5 K). This fact may suggest that for smaller steps we would measure decreasing values for $\Delta T_{vc}^a$, nevertheless we have shown that the subcritical interval increases the smaller the depth is.\par
 
At $d =$ 4 mm, for the two subsequent secondary bifurcations the fronts, in the bistable regimes, are {\it normal fronts} that connect two stable states. We have shown that these fronts are stationary for a large range of the control parameter $\Delta T_v$. As $\Delta T_v$ is increased the new unstable state will finally invade the whole cell, consequently the system has developed a global instability. This kind of behavior has also been found in catalytic reaction systems where stationary fronts appear in 1D heating with a platinum wire ~\cite{Barelko78}. Theoretically the minimum velocity of the front is achieved at the Maxwell point~\cite{Pomeau86}, so in accordance to our results we are supposed to have a ``{\it Maxwell interval}''. This interval, with a minimum and constant velocity, has already been included in some theoretical models for reaction-diffusive systems~\cite{Kramer00,Sepulchre00}. These models (for heterogeneous catalysis) show how a system with two competing diffusivities ($D_2/D_1\ll 1$) is able to sustain a quasi-stationary front for a finite interval of the control parameter with velocity $v_p\sim \sqrt{D_2}$. According to the high Prandtl number in our experiment, if we apply the previous theoretical model, the thermal diffusivity would play the role of $D_2$ and therefore, for intermediate layers this model fits the stationary front dynamics.\par  

Because of a small linear group velocity, that has been measured below the threshold to TW for a {\it FKPP front} (Fig.~\ref{fig:9}), the convective nature of the TW instability has an absolute character, therefore the threshold of the global instability is considered the same as for the convective one. 
Also, this global-convective/absolute instability to the TW pattern compares well with the theoretical work for oscillatory instabilities in the case of large $Pr$ number fluids in which the buoyancy effect is becoming important~\cite{Priede97}. This is the case for $d=$ 4 mm where $Bo_D \approx$ 1.7. On the other hand, the subcritical behavior of the system vanishes for $d \ge$ 4.5 mm, so from the codimension-2 point ($Bo_D \approx$ 2.38) and for increasing depths no fronts do exist. For a supercritical model it has been proved that the ``convective'' critical control parameter for a confined system verifies $\varepsilon_{c}=\varepsilon_a+\mathcal O(L_x^{-2})$, where $\varepsilon_a$ is the ``absolute'' critical control parameter for the corresponding extended system~\cite{Tobias98}, so for $d \ge$ 4.5 mm the approximation $\varepsilon_{c} \approx\varepsilon_a$ is valid.\par 
 
The bifurcation to ALT has also a subcritical behavior and at the threshold the pattern ALT becomes unstable at any position in the cell. The existence of a stationary front could be attributed to an inhomogeneous heating along HL. Nevertheless, the subcritical interval is about 6 K and the inhomogeneity of the temperature field between opposite sides of $L_x$ is less than 1 K. Moreover the system bifurcates supercritically for $d\ge$ 4.5 mm. The strong coupling between the different Fourier modes present for large values of $\Delta T_v$ does not allow to characterize the successive bifurcations quantitatively. 
Therefore the quantitative analysis is useful in the weak nonlinear regimes, at least for the secondary bifurcation to TW.\par 

For fluids with high Prandtl numbers these results contribute to broaden the experimental evidence about cellular patterns which undergo subcritical transitions, so modelling systems with similar behavior could be carried out in the future. Close to the threshold of the secondary bifurcation to TW, the global unstable mode by nonlinear coupling drives the dynamics through a bistable regime. The domain that has become unstable towards TW has previously crossed a mixed pattern where the resonant triad has been activated. Far from the secondary instability to the ALT pattern strong nonlinear interactions between fundamental and harmonic modes are the reason for a defect dynamics. This {\it defect mediated chaos} is the last stage of our bifurcation scenario, restricted by the available control parameter $\Delta T_v$.\par

\begin{acknowledgments}
We are grateful to L. Gil, R. Ribotta, A. Chiffaudel, N. Garnier, H. Mancini and W. Gonz\'alez--Vi\~nas for helpful discussions. This work has been partly supported by the Spanish Contract No. BFM2002-02011 and No. FIS2007-66004, and by PIUNA (University of Navarra, Spain). M.A. Miranda acknowledges financial support from the ``Asociaci\'on de amigos de la Universidad de Navarra''.
\end{acknowledgments}

\bibliography{bibart4def}

\end{document}